\PassOptionsToPackage{table,xcdraw}{xcolor}
\documentclass[sigconf,article]{acmart}

\usepackage{multirow}
\usepackage{lscape}

\AtBeginDocument{%
  }

\copyrightyear{2024}
\acmYear{2024}
\setcopyright{acmlicensed}\acmConference[FDG 2024]{Proceedings of the 19th International Conference on the Foundations of Digital Games}{May 21--24, 2024}{Worcester, MA, USA}
\acmBooktitle{Proceedings of the 19th International Conference on the Foundations of Digital Games (FDG 2024), May 21--24, 2024, Worcester, MA, USA}
\acmDOI{10.1145/3649921.3650016}
\acmISBN{979-8-4007-0955-5/24/05}

\begin{document}

\title{On the Evaluation of Procedural Level Generation Systems}

\author{Oliver Withington}
\email{o.withington@qmul.ac.uk}
\affiliation{%
  \institution{Queen Mary University of London}
  \city{London}
  \country{UK}}

\author{Mike Cook}
\email{mike@possibilityspace.org}
\affiliation{%
  \institution{Kings College London}
  \city{London}
  \country{UK}}

\author{Laurissa Tokarchuk}
\email{laurissa.tokarchuk@qmul.ac.uk}
\affiliation{%
  \institution{Queen Mary University of London}
  \city{London}
  \country{UK}}

\begin{abstract}
The evaluation of procedural content generation (PCG) systems for generating video game levels is a complex and contested topic. Ideally, the field would have access to robust, generalisable and widely accepted evaluation approaches that can be used to compare novel PCG systems to prior work, but consensus on how to evaluate novel systems is currently limited. We argue that the field can benefit from a structured analysis of how procedural level generation systems can be evaluated, and how these techniques are currently used by researchers. This analysis can then be used to both inform on the current state of affairs, and to provide data to justify changes to this practice. This work aims to provide this by first developing a novel taxonomy of PCG evaluation approaches, and then presenting the results of a survey of recent work in the field through the lens of this taxonomy. The results of this survey highlight several important weaknesses in current practice which we argue could be substantially mitigated by 1) promoting use of evaluation free system descriptions where appropriate, 2) promoting the development of diverse research frameworks, 3) promoting reuse of code and methodology wherever possible.
\end{abstract}

\begin{CCSXML}
<ccs2012>
   <concept>
       <concept_id>10010405.10010469.10010474</concept_id>
       <concept_desc>Applied computing~Media arts</concept_desc>
       <concept_significance>300</concept_significance>
       </concept>
   <concept>
       <concept_id>10003120.10003121</concept_id>
       <concept_desc>Human-centered computing~Human computer interaction (HCI)</concept_desc>
       <concept_significance>100</concept_significance>
       </concept>
 </ccs2012>
\end{CCSXML}

\ccsdesc[300]{Applied computing~Media arts}
\ccsdesc[100]{Human-centered computing~Human computer interaction (HCI)}
%%
%% Keywords. The author(s) should pick words that accurately describe
%% the work being presented. Separate the keywords with commas.
\keywords{level generation, evaluation, survey}

\maketitle

\section{Introduction}

Procedural Content Generation (PCG) for game levels and environments has developed into a highly active research over the past two decades, with multiple sub-fields concerning themselves with the generation of different types of virtual space, and with numerous different technical paradigms for generating them. It is also far from a purely academic concern, with numerous recent and highly popular game releases also relying heavily on PCG to produce their levels, ranging from titles from smaller independent studios such as Valheim \cite{irongatestudio2021} and Dwarf Fortress \cite{adams2022}, to AAA releases with large production budgets such as Diablo 4 \cite{blizardentertainment2023} and Starfield \cite{bethesdagamestudios2023}. Additionally, while game level generation is the most popular,  there are numerous other potential uses of PCG systems in games, including text and story generation, texture generation and even entire games. In this work though, when we refer to ‘PCG systems’, we are referring exclusively to generators of levels and environments.

As the level of activity and research production in this field increases, so does the importance of having generally accepted approaches and heuristics for evaluating the relative value of novel PCG approaches. High quality and widely accepted evaluation paradigms should in theory make it easier for academics and game designers to filter through the high volume of PCG research production to identify the current state-of-the-art and whether a new approach improves upon it. Other science and engineering fields have benefited substantially from such paradigms and benchmarks, with examples such as the MNIST dataset for image recognition \cite{726791} and the OpenAI Gym for training reinforcement learning agents \cite{brockman2016} having large impacts on their respective fields, and allowing for easier identification of improvements. However, PCG system evaluation has historically been a topic with limited agreement on how it should be practised despite general acceptance that it is important \cite{togelius2011,Canossa2015TowardsAP, summerville2017,liapis2020,cook2021}.

The lack of a consensus in this area is in many ways unsurprising due to the numerous challenges and nuances when evaluating PCG systems. Generated levels typically have to satisfy multiple aesthetic and functional requirements to be admissible for a given game, and the extent to which a given level satisfies these requirements can be highly subjective to evaluate. The variety of game genres that can be generated for and types of level that can be generated also problematizes the development of content agnostic techniques for evaluating generators. Any technique for evaluating and comparing generators of one specific type of digital environment, would likely be either inadmissible or need significant reworking to evaluate generators for others. Additionally, while the intention in much of PCG research is that it is useful for commercial game designers, the information on whether academic evaluative practice is aligned with their needs is limited.

To our knowledge there has never been a survey of the evaluation methods used within procedural level generator research, nor does there exist a modern taxonomy of the evaluation approaches that are available. While the practice of evaluating game level generators is often discussed in general terms, we argue that the field could benefit from a much more detailed view of current practice in the field. In this paper we provide this by first surveying a large sample of the past 5 years of research in the field, covering 86 academic publications with introduced novel systems. We then iteratively developed a taxonomy to categorise and sort these 86 works in a way that covered all evaluative practice that we encountered, based around four high-level queries about their evaluations:

\begin{itemize}
\item The method used for extracting data about the system
\item The specific features and metrics calculated
\item What, if any, is the comparison point for the system's performance
\item The game domain that the system was tested in
\end{itemize}

The intention is that this survey will provide several benefits to researchers working within this field. It highlights the breadth of possible options and paradigms for evaluating game level PCG systems, which provides researchers with a suite of potential options and access to prior works in which they have been used. It also highlights the relative frequency of use of these techniques, potentially highlighting both paradigms which are ubiquitous enough that they could be used as the basis of standardised practice, as well as under-used techniques which could be adopted more widely. As we shall see, this survey also highlights several issues and gaps within current practice which we argue could be addressed for the benefit of the field as a whole.

The primary contributions of this paper then are:

\begin{enumerate}
\item The introduction of a novel taxonomy of evaluative methods used in the field of PCG systems for generating game levels
\item Data and analysis from applying this taxonomy to a large sample of recent published work in this domain
\item Key recommendations for practitioners in this field, based on the results of the survey
\end{enumerate}

The rest of this paper is structured as follows: In Section \ref{sect_relstud} we discuss the prior studies and surveys which are most closely related to this work. In Section \ref{sect_method} we discuss the methodology we used for searching the current literature, and in Section \ref{sect_design} we introduce our taxonomy of evaluative practice and discuss its development. In Section \ref{sect_discussion} we display the results of applying this taxonomy to the surveyed works and discuss the ramifications of these results, including the potential improvements in practice that they point to. In Section \ref{sect_survlimits} we discuss the most significant limitations of the survey presented, and in the following Section \ref{sect_fut} we discuss the future trends that we think are most likely to have a substantial impact on how we evaluate PCG systems for game levels. Finally, in Section \ref{sect_conclusion} we conclude that in spite of the fields’ diversity of goals and subdomains, that there are generalisable changes in practice that could be adopted to reduce the work required to produce novel research and to increase that research’s validity. 

\section{Related Studies}\label{sect_relstud}
While to our knowledge there has never been a survey focused purely on the evaluation strategies for PCG systems nor does there exist a taxonomy for categorising them, there have been several surveys of other aspects of the field which serve as useful background and methodological inspiration. 

Arguably the most influential work which provides a survey of the field of PCG for games is 2011’s ‘Search-Based Procedural Content Generation: A Taxonomy and Survey’ from Togelius et al \cite{togelius2011}. Due to its age it has many gaps from a modern perspective, though its taxonomy of PCG systems is still highly influential. Though it does not dwell on how systems are evaluated, it does conclude its outlook section with a call for the development of methods for comparing and assessing the quality of generators. This paper formed part of the basis of 2016’s ‘Procedural Content Generation In Games’ from Shaker, Togelius and Nelson \cite{shaker2016}, which has a dedicated Chapter for PCG system evaluation. However it limits itself to discussing only two of the many possible approaches: Expressive Range Analysis (ERA) and player evaluations of content. Another older but still useful survey is that of Hendrikx et al \cite{hendrikx2013} from 2013, which provides an overview of the different types of digital artefact which can be generated using PCG techniques. While evaluation was not a focus of this work, there is another call for the development of more robust techniques for evaluating PCG systems. This call is echoed in Antonios Liapis’s ‘10 Years of the PCG Workshop: Past and Future Trends’\cite{liapis2020}, which provides an overview of work submitted to the PCG workshop, including specific insights into trends in how the works have been evaluated.

There have also been several relevant surveys of specific subdomains of PCG research. Lai et al produced a survey of Mixed-Initiative PCG systems \cite{lai2022}, and while evaluation practice was not a specific focus it is discussed for the works included. In 2023 Kutzias and Von Mammen presented a novel taxonomy and survey of papers that introduced procedural generators for digital buildings \cite{kutzias2023}. This served as direct inspiration conceptually and structurally for this work which also introduces a novel taxonomy and applies it in a survey, though the only discussion of evaluation in their survey is where it forms a component of a generative process. In 2018 Summerville et al presented a survey of Machine Learning-based PCG approaches \cite{summerville2018b}. While evaluation methods are not an explicit part of this survey, discussion of the challenges in doing so and another call for the development of comparison benchmarks forms a core part of the discussion. Finally, in 2020 Liu et al presented an overview of the artefact types generated and techniques used by Deep Learning based PCG approaches \cite{liu2021}. While evaluation was once again not a focus, it does include a specific section dedicated to the utility of deep learning techniques for evaluating content and content generators.

\section{Search Methodology}\label{sect_method}

To search for academic writing for the survey portion of this paper we used Web of Science (WoS)\footnote{webofscience.com}, a multi-disciplinary database of peer reviewed literature. As the goal of this paper is to gain insight into contemporary practice in this field, we elected to limit the search to works that had been published in the past 5 years at time of the initial search, which was the 23rd of March 2023. The query used for this survey was to search for papers whose abstracts contained both the words in the phrase ‘procedural content generation’ (not necessarily in that order or in sequence) and also the term ‘level’. Using WoS’s query language this was entered as ‘(AB=(procedural content generation)) AND AB=(level)). This query returned 138 published works. On the 30th of October 2023 we reran the same query but limited it to papers published in 2023 to capture more recent works that were published while this paper was being written, but this only returned 4 new works giving a total of 142.

By far the most popular publication venue in the sampled papers is IEEE’s Conference on Games (CoG), which was the venue for 24 of the 138 returned works. Its companion journal Transactions on Games accounted for a further 13. However dozens of other conferences and journals were represented. For the full list of works returned by these queries and how they were categorised please see our Github repository\footnote{github.com/KrellFace/PCG\_Eval\_Survey}.

\section{Taxonomy Design and Development}\label{sect_design}

The goal of this survey was to analyse contemporary practice in the evaluation of game level PCG systems. To accomplish this we needed a taxonomy of evaluation methods in this domain. However to our knowledge no such comprehensive taxonomy exists, with all of the candidates we looked at either being too partial, too broad or both. Instead we opted to develop a novel taxonomy based on what we discovered in the literature. 

Our starting point was four high level questions:
\begin{enumerate}
\item \textbf{Data Collection Method:} How is this information gathered about the system?
\item \textbf{Metrics and Features Calculated: } What information is gathered to evaluate or compare the generator's performance?
\item \textbf{Point of Comparison: } What, if anything, is the novel system compared to using this data?
\item \textbf{Game Domain: } What type of game or game-like domain is the system generating levels or environments for when it is evaluated?
\end{enumerate}

These four questions formed our four primary categories within which we developed a list of binary features which could be applied to every research paper which introduced a novel PCG system. The list of features evolved organically during the survey to include every technique encountered, as well as to differentiate between similar but distinctive approaches. When applying these features to research papers, we broadly aimed to be inclusive even if the actual use of a specified technique was clearly limited or flawed. In other words a technique did not need to be used well for a paper to be counted as using it. In this section we enumerate the features that were developed and applied, as well as detail on how these features were defined and applied.

Note that within each category the features are non-exclusive and many can appear in the same work. It was very common in the works surveyed for there to be multiple evaluative approaches used for their systems

\subsection{Initial Filter}

In order to gain useful insights from our survey we first filtered the works to only those which were both fully accessible and which introduced a novel PCG system. Of the 142 works, 3 were removed from further analysis as we could not access them through the institutional channels available to us, and 2 were removed as they were only available in languages other than English, which unfortunately is the only language that the authors of this paper speak fluently. 

Of the 137 works that remained we first needed to identify those which contained a novel PCG system for producing game levels, which required usable definitions of both PCG system and game level. Neither term is tightly defined in current literature and both can be contentious. Our aim in defining what to select for inclusion in further analysis was to adopt broad definitions of both terms which align with the general research consensus. For what counts as a PCG system the difficulties relate to the size of their output possibility spaces, and to the level of autonomy required. Systems that can only ever produce one level no matter how they are parameterised would typically not be seen as PCG systems. Similarly, systems that do not act in any way autonomously and require human input for every creative step would generally be regarded as tools and not PCG systems. Thankfully for this survey, of the papers returned from our queries, all papers that could possibly meet the criteria acted sufficiently autonomously and had large enough output spaces that their inclusion was an easy decision.

Deciding on what counts as a level or digital environment was more challenging, and we encountered several corner cases within this survey which required us to evolve our definition. The broad definition we arrived at was to include works introducing PCG systems which generated 2D or 3D spaces which are navigable by a player or NPC avatar. Two edge cases that illustrate this are the inclusion of tetris level generators such as \cite{oikawa2020}, as we judged the manoeuvring of a descending tetris pattern to count as controlling an avatar in a space, and the non inclusion of generated quizzes even if they were presented like a 2D environment if there did not appear to be a navigational element as in \cite{hooshyar2018}. We also elected to include systems which generated levels and environments even if there were no players or NPCs in the current system, so long as it was easy to conceive of how they would be included. For example, generators of digital terrain were included even if it was not possible for them to be traversed by players or NPCs because there are so many examples of games which make use of generated terrain. 

This definition is not intended to be final, and we are improving our definitions and taxonomy as we continue future work in this area. However, for this initial survey we believe the risks of including edge cases or papers that would not be described as PCG systems for levels and environments by the wider community are minimal in terms of the veracity and import of this survey’s findings. With this definition in mind, we found that 86 of the 137 introduced novel systems.

As part of this initial filter we also categorised papers that introduced a fitting system but limited themselves to being system descriptions with no evaluation. We were again inclusive in what we counted as evaluation, requiring only some level of quantitative or qualitative assessment of the results of multiple generative runs from a system. With this definition, only 5 of the 137 consisted of system descriptions only.

\subsection{Data Gathering Method}

The first category we introduce is the Data Gathering Method, by which we mean the high level approach used for gathering data about the output of a PCG system in order to evaluate it and compare it to others. Through this survey we found that there were broadly four approaches for accomplishing this, though we note this is hardly a novel finding, as the four approaches we found overlap almost completely with the three forms of individual level evaluation explained by Togelius et al in their 2011 taxonomy of search-based PCG systems \cite{togelius2011}.

\textbf{Calculated from Representation Directly -} Features that can be calculated directly from the level’s genotype or phenotype representation. This covers everything from simple counts of a certain type of block in a tile-based 2D platformer as seen in, to complex heuristics for the difficulty of a level based on the specific placement of jumps and enemies.

\textbf{Simulated Play by Agents -} Any calculation of features by giving an agent control of a surrogate player avatar would place a system evaluation here, such as the number of jump actions used to complete a level or the time taken to complete it. 

\textbf{Evaluated through Play or Observed Play by Humans -} Any gathering of data with use of human playtesters. This covers both automatically gathered data from playtraces, such as the number of player deaths, but also survey data such as asking participants to rate the enjoyability of a level on a scale.

\textbf{Evaluated through Mixed-Initiative Creation} This category is specific to Mixed-Initiative PCG systems which have a human in the loop during the design of a level mixing their input with the PCG systems decisions. Any evaluation that involved the gathering of data from humans either during or following the design of a level would come under this category. 
\subsection{Metrics and Features Calculated}

In this category we taxonomised all of the features that researchers calculated to inform the evaluations of their generators. This list evolved and expanded repeatedly during the survey process, with the overall aim of categorising every calculated feature and heuristic we discovered. The intention was that the features be specific enough that the data gathered is informative about the variety of approaches, but not so specific that the list of features expands so much that the data is hard to parse.

\textbf{Level Fitness -} A broad category that covers the use of all quantifiable heuristics which are used to quantify the quality of artefacts, most typically on a continuous scale. It can also include or incorporate other features and metrics in this list so long as the final result is a single numeric value assigned to an individual level. We also note that inclusion of a fitness function within the level generation process would not be sufficient to be included. It has to form part of the evaluation of the whole generative system’s performance.

\textbf{Solvability \& Win Rate - } Evaluation of whether or not a level can be completed, or the amount of it which can be completed. In the former case it would be a binary flag, whereas in the later it is a continuous value. In platformer level generation it is often referred to as playability \cite{summerville2016a, green2020}. Also included here are evaluations in which multiple agents are pitted against each other and their relative win rates are calculated to evaluate the level.

We note here that there is overlap between this type of feature and fitness, as some prior research has used the playability of a level as the fitness function for a search algorithm \cite{withington2020, sarkar2021}. There are also prominent works that use them separately within the same paper, especially works using Feasible-Infeasible search inspired algorithms which rely on dividing the search space for content into admissible and inadmissible spaces  \cite{liapis2015,khalifa2019,gallotta2022}. 

\textbf{Use of Validated Questionnaire -}  Data gathered from human participants using a pre-validated questionnaire such as the Game Experience Questionnaire (GEQ) and the System Usability Scale (SUS)

\textbf{Custom Questionnaire or Questions -} Data gathered from human participants using a non-validated question set. While the character of these features and how they are assessed is naturally going to be very similar to those from validated questionnaires, we were interested in the comparability of new research to prior research. Different evaluations using the same validated questionnaire are likely to be much more comparable than any using custom questionnaires, and therefore we decided to record these two types of features separately. 

\textbf{Biological Readings -} Data gathered automatically from the physical presentation of humans, including features such as eye tracking data and heart rates gathered from an ECG.

\textbf{Computational Resource Usage -} Covers any evaluation based on resource consumption rather than qualities of the artefact themselves. This includes features such as time required to generate and computing hardware required.

\textbf{Qualitative Visual Traits of a Sample -} Any discussion or comparison drawn between the aesthetic qualities of samples of generated levels would meet this category. Note, that it is not enough to just have a visualisation of a sample. It has to be discussed to some extent in the text.

\textbf{Similarity to Training Data -} Covers all generator assessments that compare a sample of generated output to its training data. This is typically relevant to machine learning based approaches which first need to be trained on exemplar levels to produce new ones, but could theoretically apply to any approach that relies on training data.

\textbf{Metric Diversity -} Includes any evaluation that involves the generation of a sample of levels, followed by a quantitative assessment of the amount of diversity present among that sample. This could be in the form of evaluating the standard deviation present in a certain feature, or more specific technique for evaluating the similarity between artefacts such as Tile Pattern KL-Divergence \cite{lucas2019}.

\textbf{Expressive Range Analysis (ERA) -} ERA is a technique for visualising a sample of output from a PCG system on a 2D plot to highlight both the diversity present and the generators tendency introduced by Smith and Whitehead \cite{smith2010}. While closely related to metric diversity, we felt that it was popular enough and distinct enough that it was important to record it separately as its focus is on extracting information from the visualisation rather than on the quantification of diversity.

\textbf{Controllability -} Covers system evaluations that were based on the system’s ability to produce levels with a variety of target characteristics or features. This is most common in systems focused on Dynamic Difficulty Adjustment (DDA) where the target characteristic is an appropriate level of difficulty for a specific player. However it can cover other types of evaluation too. 

\textbf{Performance as an Agent Curriculum -} Covers any system evaluation based on its ability to train an AI agent to perform better. For example this could be in the form of evaluating the ability of a trained agent to play more diverse levels after training with a given system, or on its ability to complete a real world task after training in a digital one produced by a PCG system. This feature is mainly specific to the subdomain of PCG research which focuses on its utility for generating appropriate training data for Reinforcement Learning (RL) agents, though it could cover any such use of PCG.

\subsection{Point of Comparison}
In this super-category we capture what, if anything, the PCG system evaluations used as their comparison point or benchmark, typically to justify a favourable assessment of their novel PCG system. In the works surveyed we found six main comparison points used in the current literature.

\textbf{System with Algorithm vs Without -} This category captures system evaluations which focus on the capabilities of a system with and without a core algorithm. This also covers evaluations which compare the output of a generative system making decisions using its core algorithm to its output when making stochastic choices.

\textbf{Alternative Parameterisations of Base System -} This category covers all system evaluations that compare the performance of the same case system but with different quantitative parameters such as different mutation rates in a genetic search based system, or different learning rates in a machine learning system.

\textbf{Alternative Algorithms Same System -}  This category covers PCG system evaluations that compare alternative named algorithms while keeping the rest of the system the same. Note that while an alternative implementation that makes completely stochastic choices is technically an alternative algorithm, these were categorised as ‘System with Algorithm vs Without’.

\textbf{System from Prior Research -} Here we cover evaluations that compare the novel system to a system from a previous research paper.

\textbf{Multiple Alternative Generators Introduced \& Contrasted -} Covers research papers which introduce multiple architecturally distinct PCG systems and compare their performance.

\textbf{Exemplar Content -} Instead of comparing to an alternative system, some PCG research evaluates their systems by comparing a sample of output to content that is presumed to be high quality, typically either content from the original game or content that is expert selected.
\subsection{Game Domain Information}
The final category covers the types of game domain that the PCG systems are designed to produce levels for and the domain they were therefore evaluated within.

\textbf{Commercial Game, Game Mod or Research Clone-} Any systems that produce levels for either a pre-existing commercial game or game mod are included here. This also covers research which involves the cloning or partial cloning of a commercial game for its test bed.

\textbf{Existing Research Platform -} Evaluations that make use of a pre-existing research platform are included here. Note that work based on research clones of games such as the Mario AI Benchmark \cite{togelius2010} also appear here so long as the clone or framework was introduced in a prior work.

\textbf{Custom Domain -} PCG systems that involve the generation of levels with a representation or game domain novel to the paper are categorised here. This is a category of exclusion, covering all works that do not either generate levels for a specific game, and that do not make use of a prior research platform or framework.

\medskip

We also opted to capture the frequency of use of two open source platforms for PCG research due to the frequency of their appearance in the work sampled. We note there are many frameworks in this field but either as a result of their infrequent use or the small size of our sample they did not appear multiple times in the works surveyed.

\textbf{Mario AI Benchmark -} Originally developed for the Mario AI competition \cite{togelius2010} it has gone on to be used in a large amount of games research, including in PCG research making it relevant here.

\textbf{GVGAI -} General Video-Game AI is an open source research platform, also originally introduced to support AI competitions \cite{perez-liebana2016}, which includes over 100 single player games and 60 multiplayer ones in a form that supports PCG research as well as other forms of games research.

\section{Results and Discussion}\label{sect_discussion}

\begin{figure}
  \centering
\includegraphics[width=\linewidth]{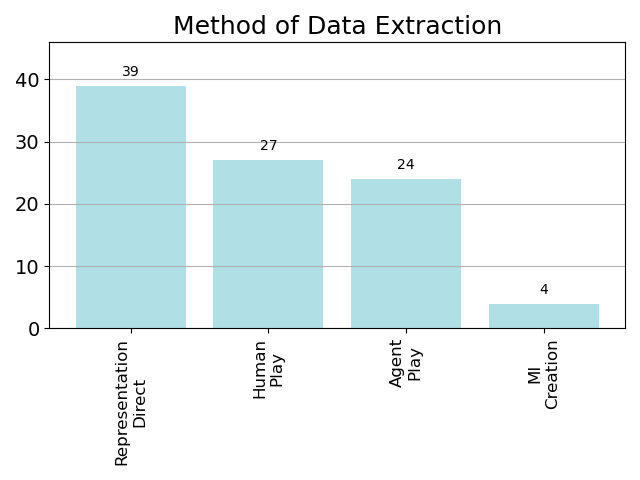}
  \caption{Histogram showing the frequency of use certain data gathering methods within the surveyed literature }
  \label{fig_evalmethods}
\end{figure}

\begin{figure}
  \centering
\includegraphics[width=\linewidth]{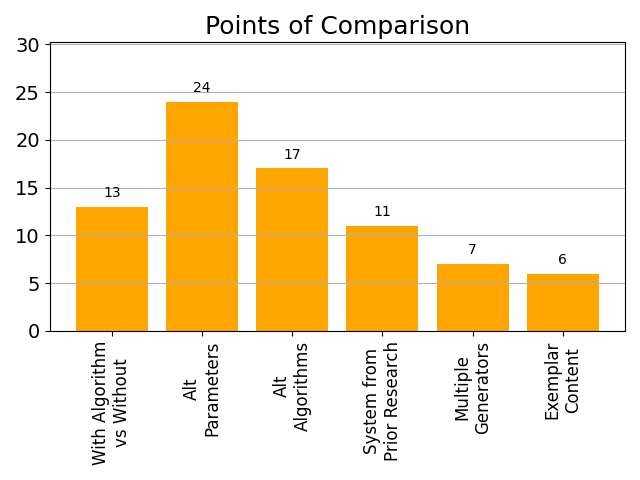}
  \caption{Histogram showing the frequency of use of different types of comparison point within the surveyed literature}
  \label{fig_comppoints}
\end{figure}

\begin{figure}
  \centering
\includegraphics[width=\linewidth]{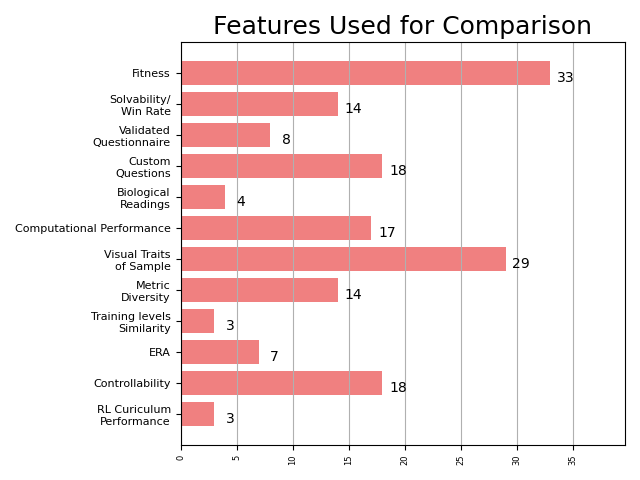}
  \caption{Histogram showing the frequency of incorporation of certain metric and heuristic types into the system evaluations within the surveyed literature}
  \label{fig_features}
\end{figure}

\begin{figure}
  \centering
\includegraphics[width=\linewidth]{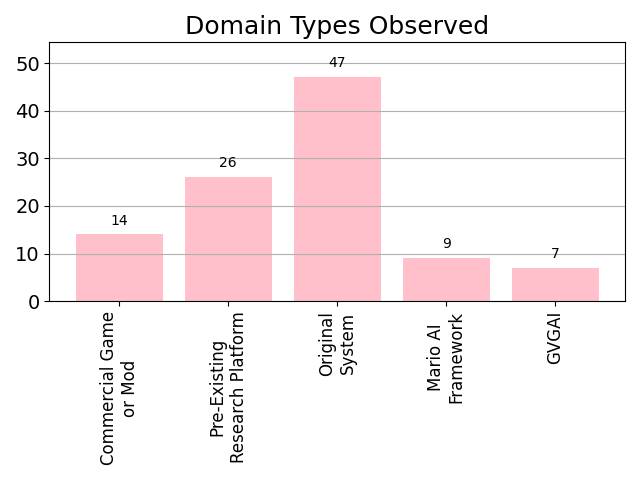}
  \caption{Histogram showing the frequency of application to certain domain types within the surveyed literature }
  \label{fig_domains}
\end{figure}

At a high level the results of this survey tell us many things about the practice of PCG research for games. Some negative and some positive. Some expected but also some surprising, at least to us. The high amount of diversity present in the field is clearly on display, especially in the breadth of metrics and heuristics used (Figure \ref{fig_features}). We were surprised by the level of enthusiasm for system evaluation in general, with only 5 of the 137 papers that introduced PCG systems consisting purely of a system description with no evaluation of it. We should note at this point however that this data can not tell us about the quality of individual works. Any of the techniques used could theoretically be appropriate and lead to robust research, and in fact we will argue later in this section that the field could be served better in some cases by not using any altogether. In the remainder of this section we will examine the individual super-categories more closely and discuss what they tell us about practice in this area and how it could be improved.

If we examine the distribution of methods used for gathering data information about the PCG systems (Figure \ref{fig_evalmethods}) we find the data indicates an enthusiasm for and commitment to system evaluation based on the frequency of use of techniques that are more time consuming and complex.  Calculation of features directly from the level representation is the most popular approach, appearing 37 times within the surveyed literature. This dominance is expected though, as calculating features in this way is typically the least work as the system designers are likely to already have direct and easy access to the representations of generated levels. We were surprised and gratified to see how common evaluation with human players was, appearing 27 times in the survey. Such evaluation requires significantly more time and labour from researchers and we would typically expect such evaluation to give us more useful data, as generated content is typically intended for human entertainment, something which is hard to predict with more abstract heuristics. We argue that this shows there is a desire to properly evaluate novel PCG systems, though we should again note that this tells us nothing about the actual quality of the research. Evaluation with human players could involve statistically insignificant sample sizes or be deeply methodologically flawed and it would still place in this category.

When we look at the distribution of comparison points used for these system evaluations (Figure \ref{fig_comppoints}) the picture is unfortunately less positive, at least if the goal is to easily identify advances in the field. The first three categories (‘With algorithm vs without’, ‘Alt parameterisation’ and ‘Alt Algorithms’) were the three most common and all three imply evaluative practice that is completely internally focused on the novel system. While the data from such evaluations can still be interesting, they cannot tell us how the system compares to alternatives that have previously been developed, and therefore are less likely to be able to inform us as to whether or not they represent an advance for the field as a whole. Evaluations that compared directly to previously developed systems were relatively rare, appearing only 11 times. A weakness of practice in this field that we argue is being indicated here is that researchers will sometimes include evaluation of their novel systems not to make a credible case that the system is a worthy inclusion in the state of the art, but to conform to expectations that such work will contain some kind of quantitative results section.

The distribution of features used (Figure \ref{fig_features}) is the most challenging category to analyse due to its diversity and scope, but we still think there are interesting features to discuss. Fitness, by which we mean quantifiable heuristics for generated level quality, is the most popular type of feature appearing 33 times within the survey. The breadth of this category makes it difficult to draw specific conclusions about the use of fitness measures or how they factor into related work outside of Game AI. However we think it highlights a similar weakness in practice to that highlighted in the comparison point data, and that is evaluation for the sake of evaluation. While fitness functions can be useful, and indeed in some types of PCG such as search-based PCG they are actively required for the system to function, they will only ever give at best a partial or at worst an actively misleading view of the quality of generated levels. Due to multiple functional and aesthetic criteria that generated levels have to satisfy, and the ultimately subjective nature of the extent to which a given artefact satisfies them, it is nearly impossible to conceive of an individual heuristic that can capture the quality of a generated level. Using such fitness functions as the basis for system evaluation risks then giving a superficial impression of quantifiable quality of a generator while in fact telling us little about whether a generator is a useful contribution to the state-of-the-art.

Elsewhere in the features used distribution we find more to be positive about. The second most popular feature is the aesthetic qualities of a sample of generated content. This view is intrinsically limited, as PCG systems can often produce many billions or more unique artefacts, meaning the largest sample of levels that could conceivably be shown in a research paper is far from statistically significant. However it is still useful for promoting a tangible understanding of the capabilities of a generator, and in fact we would argue it could often tell a reader more about a generator than an unvalidated fitness heuristic. The joint third most frequent heuristics, Custom Questionnaire and Controllability also give in our view a positive perspective on practice in the field, as both are directly linked to clearly desirable properties of PCG systems. Use of questionnaires directly ties the evaluation of the system and its output to human perceptions of its quality, and controllability ties it to a core designer need in commercial PCG i.e the capacity to produce levels with specified qualities. However we do think there is a significant under-utilisation of pre-validated questionnaires which were used only 8 times in the surveyed literature, less than half the frequency of custom questionnaires, something we discuss further in Section \ref{sect_recom}.

The survey data on the domain types observed (Figure \ref{fig_domains}) is simpler and less granular, but it highlights a substantial weakness and inefficiency in current practice, specifically the predominant use of wholly original systems for novel research. Of the works surveyed 47 of them, representing over half the sample, involved the development of original game domains or at least level representations to act as the testing domain for their PCG systems. We argue that this is to the field’s detriment for several reasons. It makes the novel PCG systems hard to compare to alternatives as comparison to any others would require reformatting one or both of them to work in a shared domain. It also is inefficient from a labour perspective, as even developing simple new game domains and content representations requires work that could be avoided by using a pre-existing benchmark or domain. For certain domains however, there may be no other options. Not all types of game level or environment have suitable research platforms, and the majority of commercial games are not easily repurposed for PCG research.

\subsection{Recommendations}\label{sect_recom}

While this data is complex and could undoubtedly benefit from wider scrutiny by the PCG research community, this data suggests to us that there are several changes that could be made to research practice which would be of general benefit.

\textbf{1. Promote Appropriate Use of System Descriptions with Limited Evaluation -} As we saw the use of some sort of evaluative practice is nearly ubiquitous in PCG research, but it is often done in a way that does not support comparisons to prior work, or in a way that improves understanding of the new approach. We also argue that for works that introduce sufficiently novel systems, comparison to prior ones might be infeasible and quantitatively evaluating them might require whole new paradigms. 

This often wasted and redundant work could be avoided by reducing the expectation that novel PCG research will contain some kind of quantitative results section, an expectation could well be due to games research having evolved out of more traditional computer science research where quantitative results are a default expectation. While robust evaluation and comparison to prior work could arguably always be desirable, developing the techniques to do so might be significantly more complex than developing the system itself, and might not even be possible at all. We argue that permitting PCG researchers to focus instead on illustrating and demonstrating the novel capacities of their new systems and not being obligated to include quantitative results could actually increase the clarity of their work. Work like this could be promoted explicitly at relevant conferences with new tracks focused on experimental new generative methods.

\textbf{2. Promote Research Framework and Benchmark Development -} The prevalence of usage of both the Mario AI Framework and GVGAI frameworks clearly demonstrates the desirability of these systems for researchers. As discussed the development and adoption of platforms like these by making it easier to exploit and compare to the prior state of the art in the field. New similarly featured frameworks for other game domains, especially 3D games which are currently underserved, would be of substantial benefit to the whole research field. 

Platform and framework development could be promoted with explicit calls at relevant conferences and journals, but it could also take the form of supporting and popularising the use of commercial games as PCG research platforms, potentially in the form of PCG competitions as has been done with Minecraft and the Generative Design in Minecraft (GDMC) competition \cite{salge2021}. This has significant advantages such as reducing the work required to develop the framework, and being able to leverage the base games popularity to increase its uptake.

\textbf{3. Promote Reuse -} As we saw in the survey data there is a significant amount of avoidable novel work that is being done to support new research in this domain. Whether it’s the development of new game systems in which to test PCG paradigms, the development of new user survey approaches or the use of novel heuristics for evaluating system performance, it all represents work that places a burden on researchers while also being hard to compare to other research. 

By promoting reuse of prior work wherever possible we can make new work more comparable to what has come previously and reduce the work required to produce novel research. We argue it also increases the likelihood that standardised evaluative practice will emerge through organically formed consensus. While the current prevalence of Super Mario-based PCG research, popularised by the Mario Framework, can be seen as a limitation of current practice, we would argue that no piece of software has done more to popularise and advance the field of PCG for games research. The wealth of prior work that it provides for reuse and the level of standardisation it provides through organic consensus has led it to being popular both for new research as well as highly innovative work from experienced researchers for well over a decade.

Working out how reuse can be promoted is less clear. In theory researchers should already be incentivised to reuse prior work as it reduces their workload and allows them to leverage prior successful research and existing consensus. However it is clear that these pressures are being outweighed by others, as we can see in survey features like the amount of new PCG work that also involves a custom domain (Figure \ref{fig_domains}). Promoting reuse of prior work can be pushed for by academic supervisors and paper reviewers, but the field could also benefit from work to assess the needs of junior researchers with a focus on barriers to producing novel PCG research.

\section{Survey Limitations} \label{sect_survlimits}
In this section we list and discuss the primary limitations of this survey and our novel taxonomy, as well as the impact we expect these limitations to have on the utility and veracity of its results.

Arguably, the biggest limitation of this survey is the extent to which it is not only non-exhaustive, but also based on a relatively small sample size of only 86 papers. There are dozens of works that introduce novel PCG systems for game levels that we are aware of which did not show in the WoS searches, and likely hundreds that we are unaware of. Furthermore WoS only indexes certain conferences and omits others, which further limits the view of the field that we are getting from this survey. However we hope that these limitations will be seen as acceptable due to both the infeasibility of a comprehensive search (searching WoS alone for published works with ‘level generation’ in the abstract produces 83,739 results at time of writing) and due to our claims about current practice in the field requiring only a representative sample of research output, not a comprehensive search.

An issue that is in some ways the inverse of the first issue is the broadness of the domain of PCG for game levels. Our scope includes very distinctive subdomains such as Mixed Initiative (MI) PCG, Dynamic Difficulty Assignment (DDA) and PCG to support RL agent training, each of which often uses distinctive evaluative techniques. Their inclusion in this survey risks reducing the clarity of the data by making some techniques appear relatively popular even though they are never used outside of their distinctive subdomain. It is unclear which subdomains it would have made sense to exclude and how one would go about defining what is included with them. Future surveys focused on evaluative practice in PCG research should consider whether it makes sense to eliminate certain subdomains with their own distinct practices from the survey. Alternatively, this limitation could potentially be avoided through further data analysis to investigate which features within categories are linked to others in different categories.

While we are satisfied with the taxonomy we presented, it is definitely worthy of further consideration and potential revision in future, especially on the Metrics and Features category and the delineations between features within it. Some of them are notably broad, especially Fitness. Fitness covers all quantifiable heuristics for quality, but quality means very different things in different pieces of research, from how aesthetically pleasing something is to how much it conforms to a desirable difficulty curve. We also did not distinguish between fitness functions which have been used in prior research and those which were introduced within the paper using them. Future surveys should consider whether to add more granularity to this category, especially due to its relative prevalence.

\section{Future Directions}\label{sect_fut}
In this section we discuss the most important future directions and trends in PCG research which are directly relevant to how we evaluate novel PCG systems

\subsection{The Gap}

For as long as PCG for games has been a field of study, researchers have noted the disconnect between PCG as researched and PCG as practised in the games industry. Shaker et al termed this “The Gap” \cite{shaker2016} . It is especially relevant to how PCG systems are evaluated, as supporting industry-based game developers in identifying which PCG approaches are most relevant and applicable for their needs could be best accomplished with evaluative practice which aligns with their requirements. 

However, in many ways academic games research has never been more closely connected with commercial game development. The popularisation of conferences such as the Roguelike Celebration\footnote{www.roguelike.club} which explicitly welcomes a mixture of both audiences, and large scale research-industrial collaborations such as those embarked upon by Unity studios\footnote{www.unity.com/academia-research}, and Nintendo and EA \cite{johannes2021} have seen robust connections made between the two. Conferences which are of central importance to the Game AI field such as Foundations of Digital Games (FDG) and the IEEE Conference on Games now actively invite academic papers on game design from commercial game designers. As the gap between academia and commercial game design appears to shrink we hope opportunities will arise to further align our evaluative practice as Game AI researchers with the needs of the game industry, at least where such alignment is the goal.

\subsection{Generative AI’s Popularity}

At time of writing, the past 3 years has seen an explosion of interest in what is now most commonly referred to as ‘Generative AI’, driven by the success of Large Language Model (LLMs) based systems like Open AI’s ChatGPT and Github’s Copilot, as well as text to image generators like Midjourney and Open AI’s DALL-E. These techniques have started to emerge as relevant in the world of PCG for game levels, with recent works like MarioGPT exploring how LLMs can be applied to the challenge of producing game content \cite{sudhakaran2023}. While it remains to be seen if these techniques will be relevant to the field in the long term given the complexity of the functional requirements for game levels, if it does become relevant or even dominant then this will have a profound effect on how we evaluate novel PCG systems. These deep learning-based approaches are extremely hard to interrogate due to their blackbox design and how contingent their output is on their training data. A sensible starting point would be to investigate the reuse of evaluative practices in other academic disciplines such as generative text and art.

\subsection{Meta Analysis of PCG Domains}

We have already discussed in this paper the diversity of subdomains and content types that can be produced with PCG systems for generating game levels, and the complications this causes for evaluating them. A potential avenue for reducing the impact of this, as well as for making PCG evaluation more robust, is the meta evaluation of PCG search spaces, an approach recently pioneered by Volz et al \cite{volz2023}. By gaining insight into how acceptable artefacts are distributed in a search space, in other words the shape of the fitness landscape for a given problem, we can make more informed statements about how difficult a given domain is to generate content for and how it compares to alternatives. Such techniques could profoundly improve evaluative practice within PCG research by supporting comparison between apparently different PCG subdomains.

\section{Conclusion}\label{sect_conclusion}
In this paper we have presented a critical overview of current practice in how novel PCG for game level systems are evaluated, through the lens of a new taxonomy of evaluative approaches constructed around four key questions about how new systems are evaluated: How is data about them extracted; What information is calculated; What are they compared to; And in what game domain is the evaluation taking place. While the picture is a complex one due to the substantial amount of diversity in the field, both in the types of content generated and the goals for it, there were trends in the data that strongly suggested changes that could be made to the field. Primarily, we argue that this field could benefit from 1) Promoting appropriate use of system description only work without the pressure to quantitatively evaluate, 2) The development of research frameworks to meet a greater diversity of needs, 3) Promoting the reuse of prior research, software and paradigms where possible, even from outside the domain of PCG for games.

\begin{acks}
This work was supported by the EPSRC Centre for Doctoral Training in Intelligent Games \& Games Intelligence (IGGI) [EP/S022325/1]. We are also grateful for the helpful and informative feedback we received from the FDG reviewers on the initial version of this work.
\end{acks}

\bibliographystyle{ACM-Reference-Format}
\bibliography{main_CameraReady}

%\appendix

%\include{appendix}

\end{document}